\documentclass[journal]{IEEEtran}
\usepackage{graphicx}
\usepackage{dcolumn}
\usepackage[mathlines]{lineno}
\usepackage{hyperref}
\usepackage{bm}
\usepackage{epstopdf}
\usepackage{amsmath}
\usepackage{epsfig}
\usepackage{bbold}
\usepackage[numbers,sort&compress]{natbib}
\usepackage{color}
\definecolor{byzantine}{rgb}{0.74, 0.2, 0.64}
\newcommand{\AF}[1]{\textcolor{black}{ #1}}
\newcommand{\BV}[1]{\textcolor{black}{ #1}}
\newcommand{\Note}[1]{\textcolor{black}{ #1}}

\newcommand{\ket}[1]{\ensuremath{\left|#1\right\rangle}}

\begin{document}
	
	\title{Creation and characterization of vector vortex modes for classical and quantum communication}

	\author{Bienvenu Ndagano, Isaac Nape, Mitchell A. Cox, Carmelo Rosales-Guzman, Andrew Forbes
		\thanks{B. Ndagano, I. Nape, C. Rosales-Guzman and A. Forbes are with the School of Physics, University of the Witwatersrand, Wits 2050, South Africa e-mail: (nibienvenu@gmail.com; isaacnape@gmail.com; carmelorosalesg@gmail.com; Andrew.Forbes@wits.ac.za).}
		\thanks{M. Cox is with the School of Electrical Engineering, University of the Witwatersrand, Wits 2050, South Africa.}
		\thanks{Manuscript received August 21, 2017; revised August 21, 2017.}}
	
	\IEEEspecialpapernotice{(Invited Paper)}

	\maketitle

\begin{abstract}
Vector vortex beams are structured states of light that are non-separable in their polarisation and spatial mode, they are eigenmodes of free-space and many fibre systems, and have the capacity to be used as a modal basis for both classical and quantum communication. Here we outline recent progress in our understanding of these modes, from their creation to their characterization and detection.  We then use these tools to study the propagation behaviour of such modes in free-space and optical fibre and show that modal cross-talk results in a decay of vector states into separable scalar modes, with a concomitant loss of information.  We present a comparison between probabilistic and deterministic detection schemes showing that the former, while ubiquitous, negates the very benefit of increased dimensionality in quantum communication while reducing signal in classical communication links.  This work provides a useful introduction to the field as well as \BV{presenting} new findings and perspectives to advance it further.
\end{abstract}

\begin{IEEEkeywords}
Vector vortex modes, quantum communication, classical communication, mode division multiplexing, geometric phase.
\end{IEEEkeywords}

\IEEEpeerreviewmaketitle

\section{Introduction}
\IEEEPARstart{S}{tructred} light, or complex light fields, has become highly topical of late, particularly as a means to realise new degrees of freedom in optical communication at both the classical and quantum regimes \cite{willner2015,Willner2017,Rubinsztein-Dunlop2017}.  This has been fuelled partly by the demand for increased bandwidth and security in communication systems, as well as by the recent advances in the ease of creation and measurement of such structured light fields \cite{Rosalesspie2017,Forbes2016}.

\BV{Employing spatial modes of light for classical communications channels has been mooted as a future technology, namely mode division multiplexing \cite{Richardson2013}, despite having been proposed more than three decades ago \cite{Berdague1982}. One constructs an orthonormal basis of spatial modes, each of which can be used as an independent communication channel upon which classical information encoding can be realised. In seminal work more than a decade ago down the corridors of the University of Glasgow,  modes carrying orbital angular momentum (OAM) have been used in a communication link \cite{Gibson2004}, and have subsequently shown, in conjunction with existing multiplexing techniques \cite{Keiser2000}, considerable improvements in data transfer in classical communication systems \cite{Wang2012, Sleiffer2012, Huang2013,Yan2014,Krenn2014,Wang2015,Ren2016,Xie2016,Willner2017,willner2015,Wang2016}.} 

\BV{At the single photon level, the use of polarization encoded qubits has recently become ubiquitous in quantum communication protocols \cite{Hubel2007,Ursin2007,Ma2012a,Herbst2015}. Most notably, they have enabled unconditionally secure cryptography protocols through quantum key distribution (QKD) over appreciable distances \cite{Jennewein2000,Poppe2004,Peng2007}. With the increasing technological prowess in the field, faster and efficient key generation together with robustness to third party attacks have become paramount issues to address. A topical approach to overcome these hurdles is through higher-dimensional QKD: increasing the dimensionality, $d$, of a QKD protocol leads to better security and higher secure key rates, with each photon carrying up to $\log_2(d)$ bits of information \cite{Bechmann2000,cerf2002,Sheridan2010}. In this regard, spatial modes carrying OAM have been used in laboratory demonstrations to show increase in photon capacity beyond what is achievable with polarisation encoded qubits, pushing the dimension to as high as $d=7$ \cite{Groblacher2006,mafu2013,mirhosseini2015}. While the list of reports on high dimensional QKD with spatial modes is not exhaustive as realizing high-dimensional quantum communication remains challenging, these seminal demonstrations are definitely promising.}

\BV{A particular form of structured light takes a so called `vector form', with spatially non-homogenous polarisation distributions, combining two well-known degrees of freedom: polarisation and spatial mode. In this form of structured light, the spatial and polarization degrees of freedom (DoFs) are coupled in a non-separable manner, reminiscent of entanglement in quantum mechanics \cite{McLaren2015}. These spatial modes are commonly known as vector modes \cite{Zhan2009} and have been used as information carriers classically in free-space \cite{Lavery2014,Zhang2016,Zhao2015,Milione2015f,Milione2015e,Li2016}, but never in fibre. In the quantum regime, vector modes have been exploited as carriers for polarisation encoded qubits in alignment-free QKD \cite{Souza2008,vallone2014}; that is because vector modes that carry OAM, also known as vector vortex modes, exhibit rotational symmetry, removing the need to align the detectors in order to reconcile the encoding and decoding bases, as would be the case in QKD with only polarization.}

Here we present an overview of state-of-the-art techniques for the generation, propagation and detection of vector vortex modes.  Using a toolkit borrowed from the quantum world, we show that entanglement measures can be used to describe such beams in a quantitative manner.   We show how they can be detected by either a filter based approach or a deterministic approach, and highlight the impact this choice has on both classical and quantum communication channels.   Using these tools we present results on the propagation of  vector vortex modes through perturbing media, namely, free-space with atmospheric turbulence and optical fibre with modal cross-talk.

\section{Basic definitions}

\begin{figure}[t]
	\centering
	\includegraphics[width=0.9\linewidth]{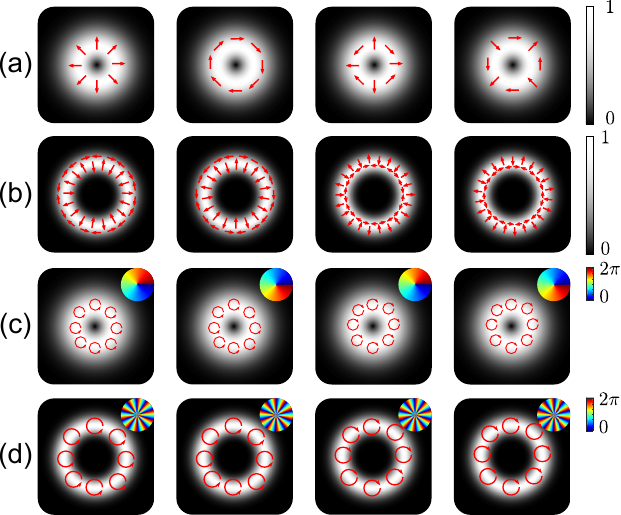}
	\caption{Vector vortex modes with (a) $\ell = \pm 1$ and (b) $\ell = \pm 10$, and their corresponding OAM scalar modes with (c) $\ell = \pm 1$ and (d) $\ell = \pm 10$. The insets show the phase profile of the scalar OAM modes.}
	\label{fig:figure1}
\end{figure}

In general, scalar OAM modes can be considered as a subset of vector vortex modes. To expand on this statement, consider an electric field $\textbf{E}$ expressed in polar coordinates $\textbf{r} = (r,\phi)$, as follows:
\begin{equation}
\textbf{E} = \cos(\theta)\exp(i\ell\phi)\hat{R} + \sin(\theta)\exp(-i\ell\phi+i\gamma)\hat{L}, \label{vectorbeam}
\end{equation}
where $\gamma$ is the intra-modal phase, $\theta$ parametrizes the amplitudes of the right- and left-circular polarisation states, respectively, labelled as $\hat{R}$ and $\hat{L}$. The functions $\exp(i\ell\phi)$ are OAM eigenfunctions of integer topological charge $\ell$. By tuning the parameter $\theta$, one controls the nature of the spatial mode in Eq.~\ref{vectorbeam}. \BV{In the instance where $\theta = \pi/4$ one obtains all vector vortex modes shown in Fig.~\ref{fig:figure1}(a) and (b) for $\ell = \pm 1$ and $\ell = \pm 10$, respectively}. Similarly, by setting $\theta = 0$ or $\pi$, one can produce either a right- or left- circularly polarised scalar OAM beam, respectively, as shown in Fig.~\ref{fig:figure1} (c) and (d).

\begin{figure}[b]
	\centering
	\includegraphics[width=\linewidth]{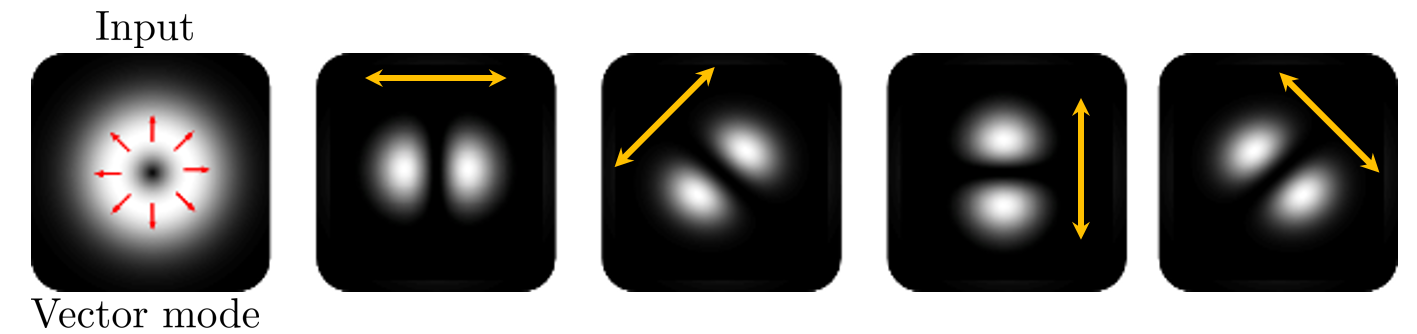}
	\caption{An input vector mode is passed through a linear polariser whose axis is shown by the yellow arrows in the figures.}
	\label{fig:figure12}
\end{figure}

\BV{It is} also evident that vector modes have intertwined degrees of freedom, space and polarisation, in a non-separable fashion; that is, the polarisation, as well as the electric field distribution, vary across the transverse plane (see Fig.~\ref{fig:figure1}). \BV{A qualitative demonstration of this relation between the polarisation and space DoFs is graphically depicted in Fig.~\ref{fig:figure12}. An input vector mode is passed through a linear polariser and the intensity measured is recorded. Observe that the two-lobe pattern obtained rotates together with the transmission axis of the polariser, indicated by the yellow arrows. One deduces that the polarisation measurement performed on a vector vortex beam, determines the spatial distribution pattern obtained, hence the non-separability.}

Vector modes have long been known to be eigenmodes of optical fibres, referred to as waveguide modes.  More recently, Milione \textit{et al.} have shown that such states of light may be described by positions on a High-order Poincar\'e sphere (HOPS) \cite{Milione2011a} that describes the total angular momentum of light, as an analogy to the familiar Poincar\'e sphere that contains only the spin angular momentum component \cite{Padgett1999}. In this pictorial representation, vector modes are mapped to points on a sphere where, the poles are \BV{circularly polarised OAM scalar modes and the equator represents the set of well-known} cylindrical vector vortex beams \cite{Zhan2009}. More than a representation, the HOPS summarises the basic recipe of how vector modes can be generated: a weighted superposition of basis scalar modes, marked with orthogonal polarisation, as given in Eq.~\ref{vectorbeam}, for all vector vortex modes.  
\begin{figure}[hb]
	\centering
	\includegraphics[width=\linewidth]{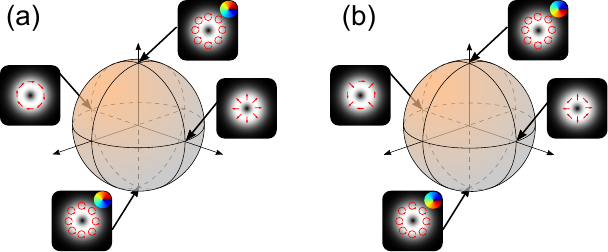}
	\caption{On the higher order Poincar\'e sphere, the poles represent the basis scalar modes from which the vector modes, located on the equator, are constructed. This is shown for basis circularly polarised OAM modes with (a) $\ell = 1$ and $\ell = -1$.}
	\label{fig:figure11}
\end{figure}

Note that in general the spatial modes would have amplitude terms as well as phase terms associated with each polarisation component. These are spatial enveloping functions that determine the electric field distribution of the vector vortex mode.  For example, the intensity profiles shown in Fig.~\ref{fig:figure1} clearly have donut profiles, which requires a radial term to the amplitude.  One could adapt Eq.~\ref{vectorbeam} to reflect this as follows:

\begin{equation}
\textbf{E} = A_{\ell}(r,\phi) \exp(i\ell\phi)\hat{R} + A_{-\ell}(r,\phi) \exp(-i\ell\phi+i\gamma)\hat{L}, \label{ampbeam}
\end{equation}

\noindent where we have dropped the weighting terms for clarity.  Here $A_{\ell}(r,\phi)$ reflects the two enveloping amplitudes (for the two signs of $\ell$).  For example, they can be tailored to produce vector OAM Bessel beams \cite{Dudley2013}, or the amplitude terms required for step-index fibre \cite{Ndagano2015} and free-space\cite{Bruning2016} propagation. The amplitude envelop is, however, often factored out of the mathematical expression of vector beams (and scalar OAM modes) because it is common to use the OAM component as the information carrier.  This is because this component can be detected in a phase-only manner, in principle without \BV{any} loss, whereas detecting the amplitude terms necessarily requires complex amplitude modulation: more complicated and lossy.  It is thus common to ignore the radial and azimuthal amplitude function when describing vector vortex beams. The vector field expressed in Eq.~\ref{vectorbeam} is however not a solution of free-space. Consequently such fields result in the excitation of additional radial eigenmodes of the medium in which they are propagating \cite{Sephton2016}. This manifests in the excitation of undesired radial modes with a large energy component \cite{Sephton2016}. 

Recently it has become commonplace to formulate the above in a quantum language, reflecting the fact that such modes of light are non-separable in their two degrees of freedom, akin to quantum entanglement.  This so-called classical entanglement \cite{Spreeuw1998,Pereira2014,Toppel2014a,Guzman-Silva2015a,Karimi,Karimi2010,DAmbrosio2013} is highly controversial but has also found some practical advantages over separable (scalar) states of light \cite{Ndagano2017,Bromberg2009,Keil2010,Keil2011,Lerman2010,Michihata2009,Berg-Johansen2015}.  Using Dirac's notation, we can denote a four dimensional space spanned by four scalar OAM modes, $\{ \ket{\ell,L},\ket{-\ell,L}, \ket{\ell,R},\ket{-\ell,R} \} $, to form the cylindrical vector vortex modes as

\begin{eqnarray}
\ket{\mbox{TE}} &=& \frac{1}{\sqrt{2}}(\ket{\ell,R} + \ket{-\ell,L}), \\
\ket{\mbox{TM}} &=& \frac{1}{\sqrt{2}}(\ket{\ell,R} - \ket{-\ell,L}), \\
\ket{\mbox{HE}}_e &=& \frac{1}{\sqrt{2}}(\ket{\ell,L} + \ket{-\ell,R}), \\
\ket{\mbox{HE}}_o &=& \frac{1}{\sqrt{2}}(\ket{\ell,L} + \ket{-\ell,R}), 
\end{eqnarray}

The interest in these vector states as a basis for information transfer resides in the fact that there are an infinite number of such four dimensional spaces, one for each value of $|\ell|$, i.e., an infinite number of HOPS to exploit.  \AF{For each HOPS we can assign the elements 00, 01, 10, and 11 to the TE, TM, HE$_e$ and HE$_o$ modes as our information carriers.}  Fortunately, the means to create and detect such states on the HOPS have also recently been developed.

\section{Generation of vector vortex modes}


Vector vortex beams have long been created internal to lasers by exploiting gain competition in birefringent laser crystals \cite{Forbes2016} and more recently with geometric phase \cite{Naidoo2016} and with structured materials in on-chip solutions \cite{Iwahashi2011,Cai2012,Naidoo2015,Shu2016,Miao2016}.  While we acknowledge that such custom solutions exist, they are not common to most laboratories.  Here we consider more standard and accessible approaches to the creation of such modes.

From Eq.~\ref{vectorbeam}, one deduces that in order to generate a vector beam though dynamic phase manipulation requires a device that can independently modulate the two polarisation components. This can be achieved using a spatial light modulator (SLM); the liquid crystals present in the display of an SLM can be individually addressed in order to spatially control the refractive index distribution of the digital hologram used to modulate the incident beam (see Refs. \cite{Forbes2016} or a comprehensive review on beam shaping with SLMs). The birefringent liquid crystals only allow phase modulation of one polarisation component and so special tricks are needed to overcome this for vector beam generation \cite{Rosalesspie2017}. \BV{Note that it has been shown that spatial modes can also be generated with digital micromirrors \cite{Mirhosseini2013,Gong2014,Ren2015}}.

\begin{figure}[t]
	\centering
	\includegraphics[width=\linewidth]{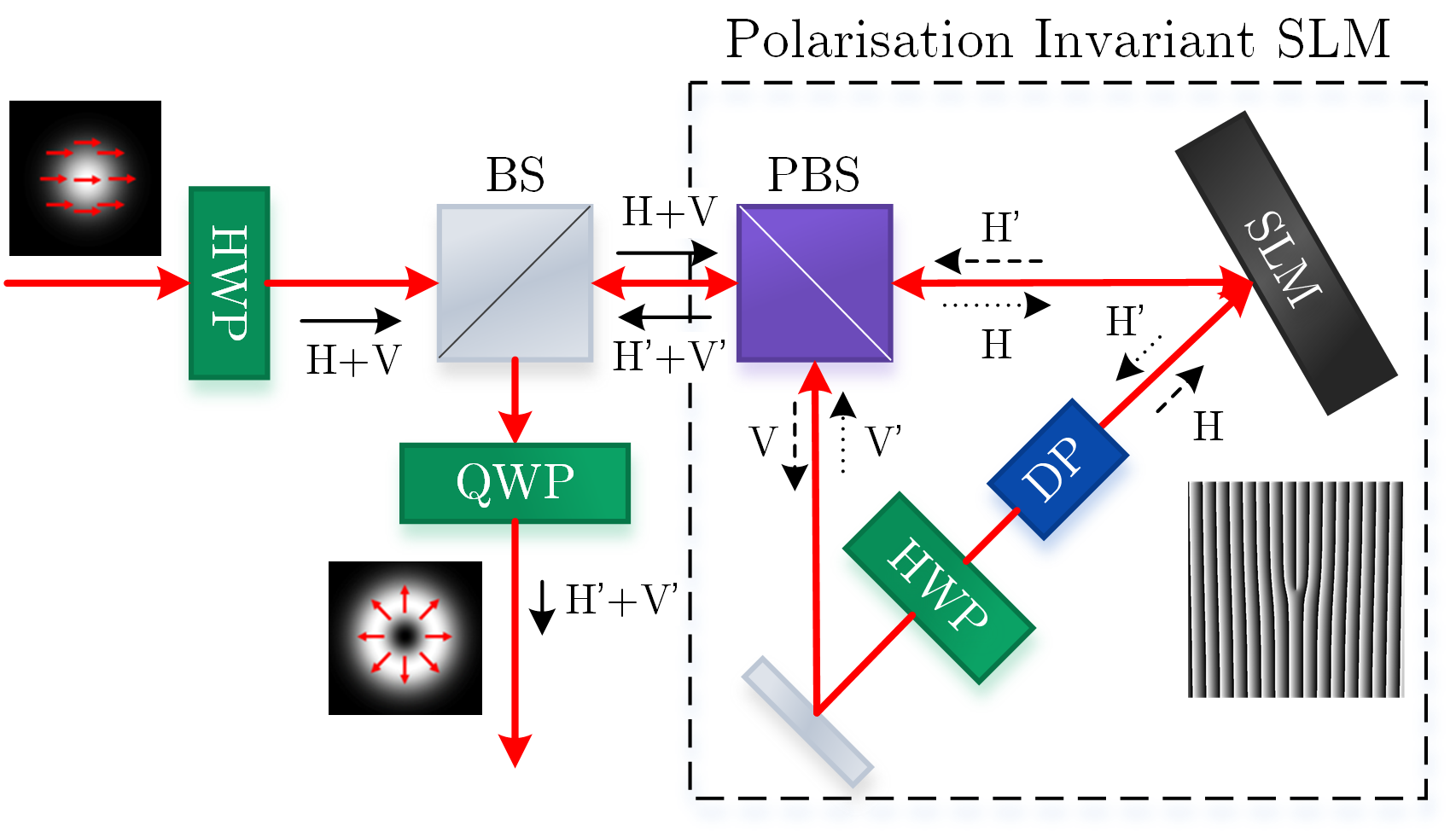}
	\caption{An incident polarised laser beam is spatially divided in horizontal and vertical polarisation components using a polarising beam splitter (PBS), each of which is modulated by a single hologram encoded on an SLM, placed inside a Sagnac interferometer. The dove prism (DP) ensured that the two OAM modes produced have opposite charge $\ell$. BS: 50:50 beam splitter; HWP: half-wave plate; QWP: quarter-wave plate. H, H', V and V' denote the polarisation paths before and after the SLM.  }
	\label{fig:figure2}
\end{figure}

Early use of SLMs to tailor vector vortex beams involved placing them inside an interferometer, as first demonstrated by Neil \textit{et al.} \cite{Neil2002}. By splitting the beam into orthogonal polarisations, one can align the polarisation axes of the two beams to match the SLM, independently modulate the two beams before recombination at a later stage, as shown in Fig.~\ref{fig:figure2}. Due to the difference in the number of reflections after the beam interacting with the hologram encoded on the SLM, each polarisation component acquires oppositely OAM values. This approach was later generalised in \cite{Maurer2007}. Other variants of the generation of vector modes with SLMs involve a double pass on each of the two different hologram \cite{Moreno2012}. In between each pass, the polarisation is rotated to allow modulation of only one polarisation component.

Spatial light modulators have the inconvenience of high losses, since only the first diffracted order contains the desired modulation. The independent modulation could be realised with the use of spiral phase plates. However, they are non-trivial to manufacture for optical wavelengths and are offer no versatility in terms of operating wavelength and spatial mode produced (unlike an SLM). A novel approach to beam shaping of vector beams makes use geometric phase rather than dynamic phase.  First introduced by Pancharatnam and later generalized by Berry, the geometric phase naturally arises when a system, in an initial state on a sphere is moved along a closed path, back to its initial position \cite{Pancharatnam1956,Berry1984}. Galvez \textit{et al.} have shown that the same geometric phase also arises from similar transformations on the OAM Bloch sphere \cite{Galvez2003, Padgett1999}.

The geometric phase equally arises as a result of birefringence. As a beam propagates through an anisotropic medium, the orthogonal polarisation states experience different refractive indices, resulting in different optical phases on each polarisation state. The nature of the phase acquired depends on the geometry of the birefringence, hence a geometric phase. A recent example of this is the so-called \textit{q}-plate, which has a locally varying birefringence resulting in spin-orbit coupling \cite{Marrucci2006,Marrucci2011}. The transformation of a \textit{q}-plate can be summarised as

\begin{eqnarray}
\exp(i\ell\phi)\hat{L}&\xrightarrow{q\text{-plate}}&\exp(i(\ell+2q)\phi)\hat{R} \label{qplate1}\\
\exp(i\ell\phi)\hat{R}&\xrightarrow{q\text{-plate}}&\exp(i(\ell-2q)\phi)\hat{L} \label{qplate2}
\end{eqnarray}
where \textit{q} is the charge of the \textit{q}-plate. \BV{By this approach all states shown on the HOPS in Fig.~\ref{fig:figure11} can be produced, as shown in the Table below.} 
\begin{table}[h]
	\centering
	\caption{Generation of scalar and vector vortex modes through geometric phase control using a \textit{q}-plate. The input mode in all cases is a Gaussian beam with OAM charge $\ell = 0$. }
\begin{tabular}{|c|c|c|}
	\hline 
	Input polarisation & q-plate charge & output mode \\ 
	\hline 
	$\hat{R}$ & 1/2 & $\exp(-i\phi)\hat{L}$ \\ 
	\hline 
	$\hat{L}$ & 1/2 & $\exp(i\phi)\hat{R}$ \\ 
	\hline 
	$\hat{R}$ & -1/2 & $\exp(i\phi)\hat{L}$ \\ 
	\hline 
	$\hat{L}$ & -1/2 & $\exp(-i\phi)\hat{R}$ \\ 
	\hline 
	$\hat{H}$ & 1/2 & TM $(\ell = 1)$ \\ 
	\hline 
	$\hat{V}$ & 1/2 & TE $(\ell = 1)$ \\ 
	\hline 
	$\hat{H}$ & -1/2 & HE$_e\ (\ell = 1)$ \\ 
	\hline 
	$\hat{V}$ & -1/2 & HE$_o\ (\ell = 1)$ \\ 
	\hline 
\end{tabular} 
\end{table}
\BV{It is worth noting the geometric phase transformation outlined in Eqs.~\ref{qplate1} and \ref{qplate2}, have been implemented experimentally using sub-wavelength gratings \cite{Bomzon2001,Bomzon2002,Biener2002,Hasman2003,Niv2005} and metamaterials \cite{Zhao2013,Yi2014,Yue2016}}. What remains then is to determine which mode you have produced, and to what purity.

\section{Detection of vector vortex beams}
The determination of the vectorial nature of an optical field has typically been qualitative or at best an average degree of polarisation across the field. Due to the spatially variant polarisation map, vector vortex beams produce intensity patterns that vary when passed through a rotating polariser \cite{Zhan2009}, and one routinely sees such plots as a confirmation of the vector nature of the light under study. Such a measurement is however not adequate to determine quantitatively the quality of the vector beam produced. Nowadays, there are two approaches to quantitatively measure the quality of a vector vortex beam. Before we get into the details of these techniques, it is important to understand their origins.
\begin{figure}[t]
	\centering
	\includegraphics[width=\linewidth]{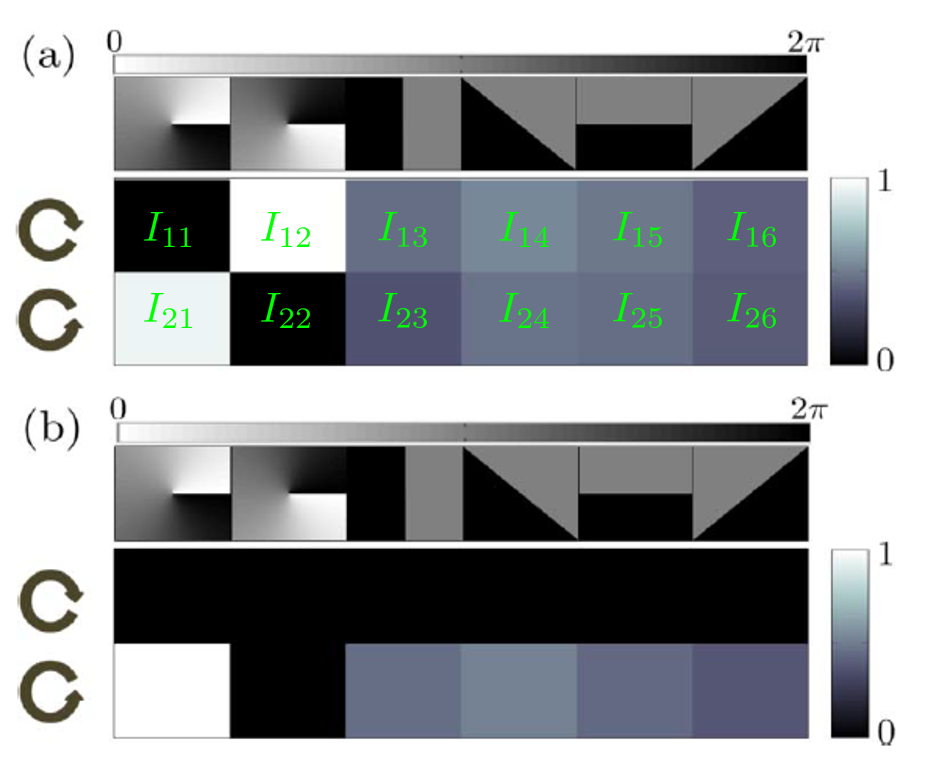}
	\caption{Normalized projections of the left- and right-circular polarisation states, onto OAM eigenstates of the Pauli matrices for (a) a vector mode and (b) a scalar mode.}
	\label{fig:figure4}
\end{figure}
\subsection{The vector quality factor}
The first measure quantifies non-separability. From the expression of a vector beam in Eq.\ref{vectorbeam} we see that depending on the value of $\theta$, the superposition is either non-factorizable or factorizable in terms of the individual degrees of freedom.  For example, for $\theta = \pi/4$ the electric field $\textbf{E}$ cannot be expressed as a product function of the following kind: $\textbf{E}=space\times polarisation$. The space and polarisation DoFs are thus said to be non-separable. We can define a new beam quality factor for vector beams, a `vector quality factor' \cite{Ndagano2016,McLaren2015}. The measure finds its origins in the quantum world, where measures of entanglement quantifies non-separability between multi-photon states. With respect to a vector mode defined as in Eq.~\ref{vectorbeam}, the vector quality factor (VQF) is defined as 
\begin{equation}
\text{VQF} = \text{Re}\left(\sqrt{1-s^2}\right) = 2|\cos(\theta)\sin(\theta)|, \label{VQF}
\end{equation}
where $s^2$ is the length of the Bloch vector, defined with respect to expectation values of the Pauli matrices $\sigma_i$:
\begin{equation}
s^2 = \sum_{i=1}^{3}\langle \sigma_i \rangle^2
\end{equation}
\begin{figure}[b]
	\centering
	\includegraphics[width=\linewidth]{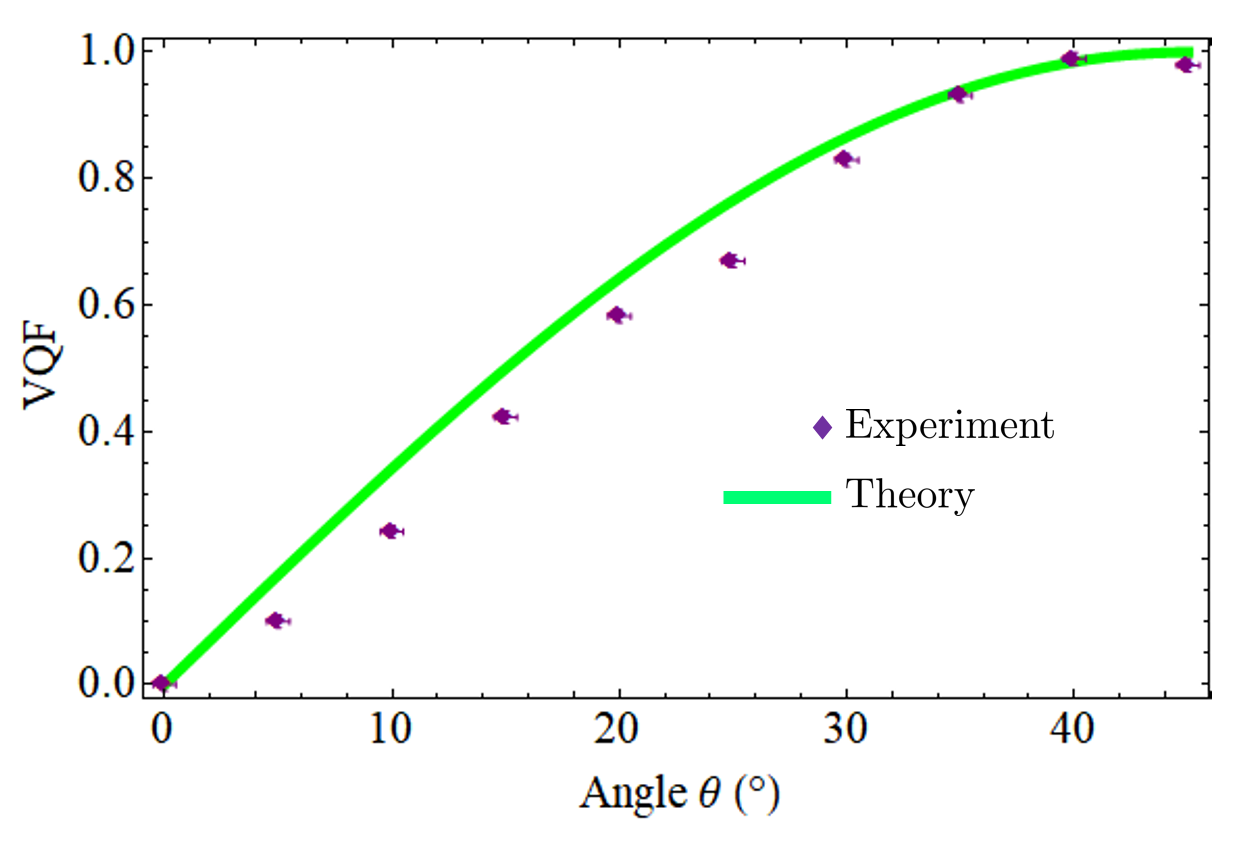}
	\caption{Experimental and theoretical variation of the VQF with respect to the amplitude parameter $\theta$}
	\label{fig:figure5}
\end{figure}
The expectation values $\langle \sigma_i \rangle$ are obtained by performing an optical projection of the vector beam onto the eigenvectors of the matrices $\sigma_i$. There is of course the issue of identifying the DoF in which the projections are realised. Here, the choice is arbitrary; one can either project each OAM eigenmode onto polarisation eigenstate or vice-versa. The latter is preferred as it can be automated \BV{using either an SLM or a digital micromirror, allowing for simultaneous measurements of all expectation values through spatial mode demultiplexing}. The expectation values $\langle \sigma_i \rangle$ are computed as follows:
\begin{eqnarray}
\langle\sigma_1\rangle &=& (I_{13} + I_{23}) - (I_{15} + I_{25}),\\
\langle\sigma_2\rangle &=& (I_{14} + I_{24}) - (I_{16} + I_{26}),\\
\langle\sigma_3\rangle &=& (I_{11} + I_{21}) - (I_{12} + I_{22}).
\end{eqnarray}

The layout of the projections is shown in Fig~\ref{fig:figure4}. The grayscale images on the top represent the phase holograms encoded on the SLM to realize the spatial projections.

By measuring the expectation values $\langle \sigma_i \rangle$ as a function of the parameters $\theta$, one obtains a variation of the VQF from 0 to 1 for $0\leq\theta\leq\pi/4$, as shown on Fig.~\ref{fig:figure5}. Note that varying $\theta$ from $0$ to $\pi/4$ is equivalent to a motion on the HOPS, from the pole to the equator where the vector vortex modes with maximum non-separability are located.

\subsection{Deterministic and filter-based detections}
\begin{figure}[t]
	\centering
	\includegraphics[width=\linewidth]{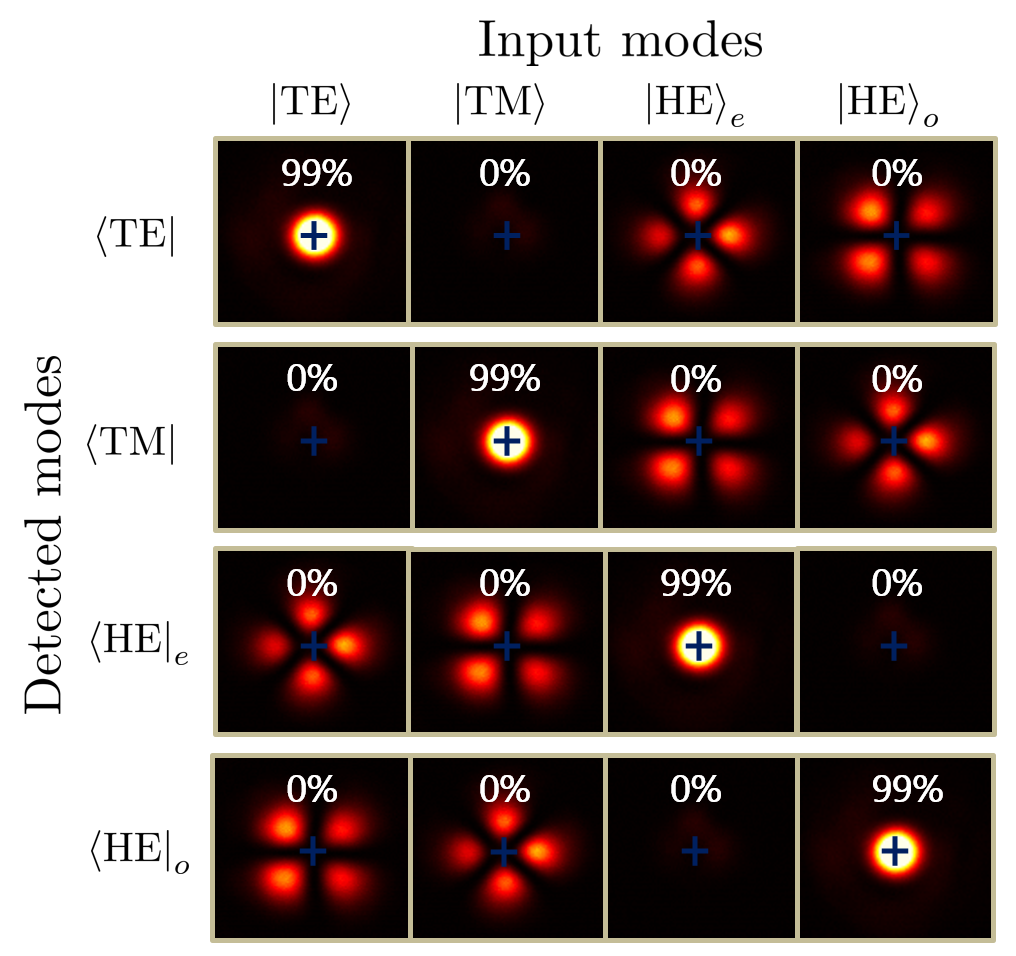}
	\caption{Experimental Modal decomposition of vector vortex modes in the $\ell = \pm 1$ subspace.}
	\label{fig:figure13}
\end{figure}

The second quantitative measure of `vectorness' is the traditional modal decomposition, extensively demonstrated in the case of scalar beams \cite{Kaiser2009,Schulze2012,Litvin2012,Schulze2013a}. \BV{It is however useful to go over the technique in the case of vector vortex beam. Note that the following example can be generalised to other forms of vector and scalar modes. The aim of vector modal decomposition is to express an arbitrary optical field $\textbf{U}$ as a weighted superposition of basis vector modes $\psi_{\ell}$: $\textbf{U} = \sum_{{\ell}}c_{\ell}\psi_{\ell}$, where $c_{\ell}$ are complex coefficient, appropriately normalized; that is $\sum_{{\ell}}|c_{\ell}|^2 = 1$. in the case where vector vortex modes are used as basis, we can write the basis states $\psi_{\ell}$ as follows:
\begin{equation}
\psi^{\pm}_{\ell} = \frac{1}{\sqrt{2}}\left(\exp(i\ell\phi)\hat{R} \pm \exp(-i\ell\phi)\hat{L}\right).
\end{equation}
Let \textbf{U} be a vector vortex mode in this basis that is used to carry a given stream of information. We write \textbf{U} as follows:
\begin{equation}
\textbf{U} = \frac{1}{\sqrt{2}}\left(\exp(im\phi)\hat{R} + \exp(-im\phi)\hat{L}\right).
\end{equation}
The coefficients $|c_{\ell}^{\pm}|^2$, can be associated with physical detectors, and are given as follows
\begin{eqnarray}
|c_{\ell}^{\pm}|^2 &=& \frac{1}{2\pi}\sum_{{\ell}}\left|\int_{\mathbb{R}^2} d\mathbb{R}^2\ \psi^{\pm^*}_{\ell}\textbf{U}\right|^2,\\
&=&\frac{1}{4\pi}\sum_{{\ell}} \left|\int_{\mathbb{R}^2}d\mathbb{R}^2\  e^{i(m-\ell)\phi} \pm e^{-i(m-\ell)\phi}\right|^2,\\
&=&\frac{1}{2}\sum_{{\ell}} \left|\delta_{m,\ell} \pm \delta_{m,\ell}\right|^2.
\end{eqnarray}
One then deduce the following result
\[|c_{\ell}^{-}|^2 = 0\ \text{and }|c_{\ell}^{+}|^2 \begin{cases} 
1 & \ell = m \\
0 & \ell \neq m 
\end{cases}
\]
The implication is that only one vector mode detector, among the multitude one can construct in this basis, will register a signal from the transmission on \textbf{U}. A graphical example of such a detection is shown in Fig.~\ref{fig:figure13}. Within a four-dimensional basis set of vector modes with OAM charge $\ell = 1$. each element is decomposed with respect to itself and the others. The cross-talk, that is the amount of power measured in a given mode $i$ for an input mode $j$, is numerically shown for each input mode.}

\begin{figure}[t]
	\centering
	\includegraphics[width=\linewidth]{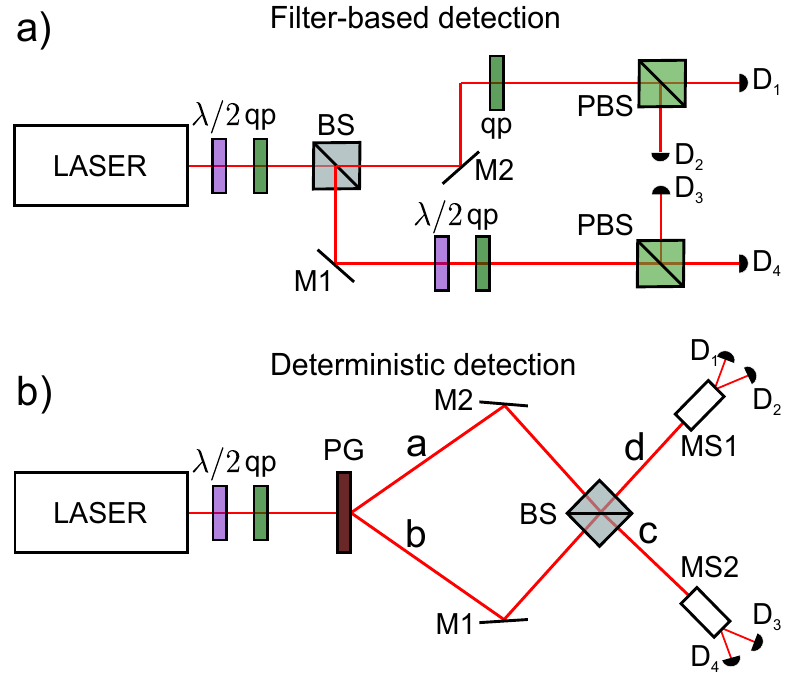}
	\caption{A vector vortex beam is generated by transforming a linearly polarised Gaussian laser beam with a half-wave plate and a \textit{q}-plate. (a) The vector beam is sent along two path, each having a vector mode filter based on geometric phase. (b) The generated vector beam is passed through a Mach-Zehnder interferometer that resolves intra-modal phases. The two outputs of the interferometer are sent to OAM mode sorters that resolves the OAM content.}
	\label{fig:figure6}
\end{figure}

This vector mode decomposition was first introduced by Milione \textit{et al.} in the context of classical vector mode multiplexing through optical fibres \cite{Milione2015f}.  As depicted in Fig.~\ref{fig:figure6}(a) a vector beam generated with geometric phase optics is set along two paths. In each of the paths, a \textit{q}-plate, together with a polarising beam splitter act as a filter for vector modes. This method however comes with an inherent 50\% loss in intensity (photon number).
Note that the two arms contain slightly different filters: one has a half-wave plate while the other does not. It can be shown by matrix algebra that the half-wave plate changes the sign of the charge of the $q$-plate. The implication is that each arm is set to only detect a maximum of two vector vortex modes, the other two giving identical signals. At the classical level, demultiplexing vector modes shown in Fig.~\ref{fig:figure11} using the setup in Fig.~\ref{fig:figure6}(a) would result in lower signal-to-noise ratio due to the 50\% loss at the beam-splitter. In the context of quantum key distribution, such a loss would result in lower sifting rates and effective key rate, negating the benefit of the higher-dimensional space provided by spatial modes. It is in light of the above that we introduce a novel method based on mode-to-space mapping; that is, a spatial mode is mapped to a set of spatial coordinate with, in principle, unit probability.
\begin{figure}[b]
	\centering
	\includegraphics[width=\linewidth]{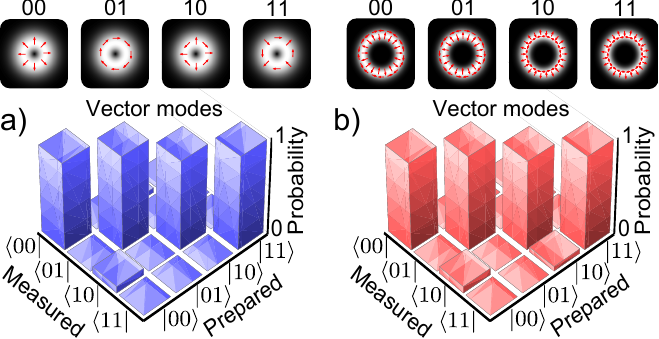}
	\caption{Crosstalk measurement for vector modes in the subspaces (a) $\ell = \pm 1$ and (b) $\ell = \pm 10$.}
	\label{fig:figure8}
\end{figure}

Consider a vector mode defined as follows
\begin{equation}
\textbf{E} = \frac{1}{\sqrt{2}}\left(\exp(i\ell\phi)\hat{R} + \exp(-i\ell\phi+i\gamma)\hat{L}\right). \label{vectorbeam2}
\end{equation}
The sorting of the different vector modes is achieved through a combination of geometric phase control and multi-path interference as shown in Fig.~\ref{fig:figure6}(b).  First, a polarisation grating based on geometric phase \cite{Li2012} acts as a beam splitter for left- and right-circularly polarised photons, creating two paths

\begin{equation}
\textbf{E} = \frac{1}{\sqrt{2}}\left(\exp(i\ell\phi)\hat{R}_a + \exp(-i\ell\phi+i\gamma)\hat{L}_b\right),
\label{eq: vector mode2}
\end{equation}
where the subscript $a$ and $b$ refer to the polarisation-marked paths. The photon paths $a$ and $b$ are interfered at a 50:50 BS. The optical phase difference between the path is set to $\pi/2$, resulting in the following state after the BS:
\begin{equation}
\textbf{E}' = \left(\frac{1-e^{i\gamma}}{2}\right)\exp(i\ell\phi)_c + i\left(\frac{1+e^{i\gamma}}{2}\right)\exp(-i\ell\phi)_d. \label{eq: path interference}
\end{equation}
where the indices $c$ and $d$ label the output ports of the BS. Note that the polarisation of the two paths is automatically reconciled in each of the output ports of the beam splitter due to the difference of parity in the number of reflections for each input arm. Hence, we drop the polarisation vectors in each output arm.  Also note that at this point it is not necessary to retain the polarisation kets in the expression of the photon state since the polarisation information is contained in the path.
\begin{figure}[t]
	\centering
	\includegraphics[width=\linewidth]{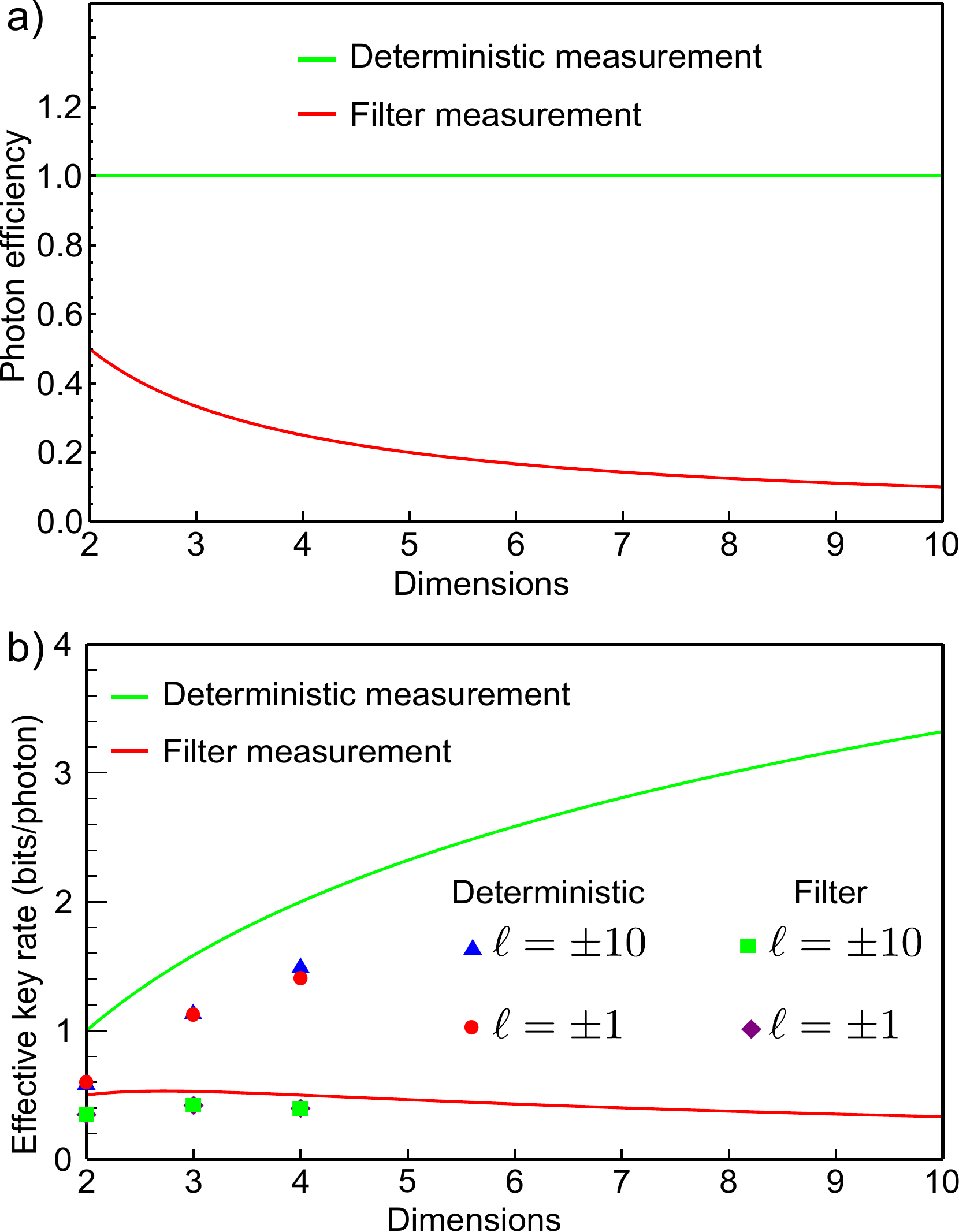}
	\caption{(a) Photon efficiency as a function of dimension for a deterministic and filter-based measurement scheme. (b) Effective key rate as a function of dimension shown experimentally (data points) for two measurement approaches, namely, filtering and deterministic detection. Data points show measurements in the subspaces of $\ell = \pm 1$ and $\ell = \pm 10$ for $d$ = 2,3 and 4. The measurement fidelities for $\ell = \pm 1$ and $\ell = \pm 10$ were measured to be 0.96 and 0.97, respectively. The solid curves are theoretical bounds on the effective key rate assuming a fidelity of 1.}
	\label{fig:figure7}
\end{figure}
The measurement system is completed by passing each of the outputs in $c$ and $d$ through a mode sorter and collecting the photons using 4 fibres coupled to avalanche photodiodes.  The mode sorters are refractive (lossless) aspheres that map OAM to position \cite{Berkhout2010a, Fickler2014a, Lavery2013, Dudley2013}. The mapping is such that 

\begin{eqnarray}
\gamma = 0 &\rightarrow& \textbf{E}' = i \exp(-i\ell\phi)_d, \\ 
\gamma = \pi &\rightarrow& \textbf{E}' = - \exp(i\ell\phi)_c, 
\label{eq:mapping}
\end{eqnarray}
In each output arm $c$ and $d$, the OAM mode sorters maps $\pm \ell$ OAM state to distinct spatial positions. In this way, all the vector modes defined as in Eq.~\ref{vectorbeam2} can be mapped to independent positions, thus achieving a complete sorting of vector modes with unit efficiency. This holds at both the classical and quantum levels. Graphical representation of the cross-talk measured for vector modes of the $|\ell| = 1$ and $|\ell| = 10$ subspaces is shown in Fig.~\ref{fig:figure8}

The advantage of our approach is graphically depicted in Fig.~\ref{fig:figure7}(a), where we plot, as a function of the dimension, the photon efficiency of the detection: defined as $1 - S$ where $S$ is the fraction of photons whose information is lost due to measurement approach. Since a filter can only probe one spatial mode at the time, the measurement of a $d$-dimensional state will return a positive detection with an average probability $\leq 1/d$. In two dimensions, two modes can be detected at once by inferred from the state of a single detectors; given that the probability of a detection in detector A is $p_A$, the probability of detection in detector B is thus $1-p_A$. 

Our system on the other hand does not suffer any dimension dependent loss. This would make it particularly beneficial to both classical communication through mode-division multiplexing with higher signal-to-noise ratio, as well as high efficiency quantum communication in the form of high-dimensional quantum key distribution. The latter is graphically illustrated in Fig.~\ref{fig:figure7}(b) where we plotted theoretical and experimental data of the \Note{photon information capacity} as a function of the dimension. By photon information capacity we mean the $d$-dimensional effective key rate as defined in \cite{Ferenczi2012}, taking into account detection efficiency. Note that the filter-based detection has key rates well below the qubit (\textit{d}=2) maximum of 1 bit per photon.

\section{Propagation through perturbing media}
When propagating through various communication channels (free-space and fibres), vector vortex modes are affected by the imperfections of the medium, for example, turbulence in free-space and impurities and stress in optical fibres. This results in intermodal cross-talk and interference which affects the spatial mode of interest. Consider for example the propagation of a radially polarised vector vortex mode through turbulence:
\begin{equation}
\textbf{E} = \frac{1}{\sqrt{2}}\left(\exp(i\ell\phi)\hat{R} + \exp(-i\ell\phi)\hat{L}\right). \label{vectorbeam3}
\end{equation}
The atmosphere is largely non-birefringent. As such, the polarisation is unaffected during propagation. The spatial degree of freedom however is highly susceptible to atmospheric turbulence, causing a scattering among OAM states \cite{Paterson2005,Malik2012,Rodenburg2012,Goyal2016,Chen2016}. Let us consider vector modes of a given OAM subspace $\ell$. After turbulence, the final vector modes state can be expressed as follows
\begin{equation}
\begin{split}
\textbf{E}'(\xi',\xi) =& \frac{\cos(\xi)\exp(i\ell\phi) + \sin(\xi)\exp(-i\ell\phi)}{\sqrt{2}}\hat{R}\\
& + \frac{\cos(\xi')\exp(-i\ell\phi) + \sin(\xi')\exp(i\ell\phi)}{\sqrt{2}}\hat{L},
\end{split}
\end{equation}
where $\xi$ and $\xi'$ parametrize the scattering amplitude between OAM modes in the $\ell$ subspace, with $0\leq\xi,\xi'\leq\pi/4$. In the absence of turbulence, $\xi=\xi'=0$ and one recovers the state in Eq.~\ref{vectorbeam3}. In infinitely strong turbulence, $\xi=\xi'=\pi/4$; that is, all OAM states have equal probability and the information on the initial state is `erased'.  Given the structure of the atmosphere, it can be considered a non-chiral medium; i.e., the $\pm \ell$ OAM eigenstates should experience similar scattering when going during propagation. Thus, it is reasonable to assume $\xi = \xi'$. Borrowing tools from quantum mechanics, the VQF of the state $\textbf{E}'$ can be expressed in the same fashion as the concurrence for entangled state \cite{Wootters2001}:
\begin{equation}
\text{VQF}(\textbf{E}') = |\cos^2(\xi)-\sin^2(\xi)| = |\cos(2\xi)|.
\end{equation} 
For $\xi = 0$, we have $\textbf{E}'(0) = \textbf{E}$; that is, in the absence of turbulence, the final state is a pure vector mode with VQF = 1. In the case of infinitely strong turbulence, $\xi = \pi/4$ and after some simple algebra, one can show that the final state \textbf{E}' is expressed as follows
\begin{equation}
\textbf{E}'(\xi) = \frac{1}{\sqrt{2}}\left(\exp(i\ell\phi) + \exp(-i\ell\phi)\right)\left(\hat{R}+\hat{L}\right) = \cos(\ell\phi)\hat{H}, \label{vectorbeam4}
\end{equation}
where $\hat{H} = \left(\hat{R}+\hat{L}\right)/\sqrt{2}$ is the horizontal polarisation state. By definition, $\textbf{E}'(\pi/4)$ is a scalar mode and therefore, a separable state. As such, $\text{VQF}(\textbf{E}'(\pi/4)) = 0$.  A similar analysis holds for optical fibre since the OAM modes are degenerate and thus couple strongly together.
\begin{figure}[t]
	\centering
	\includegraphics[width=\linewidth]{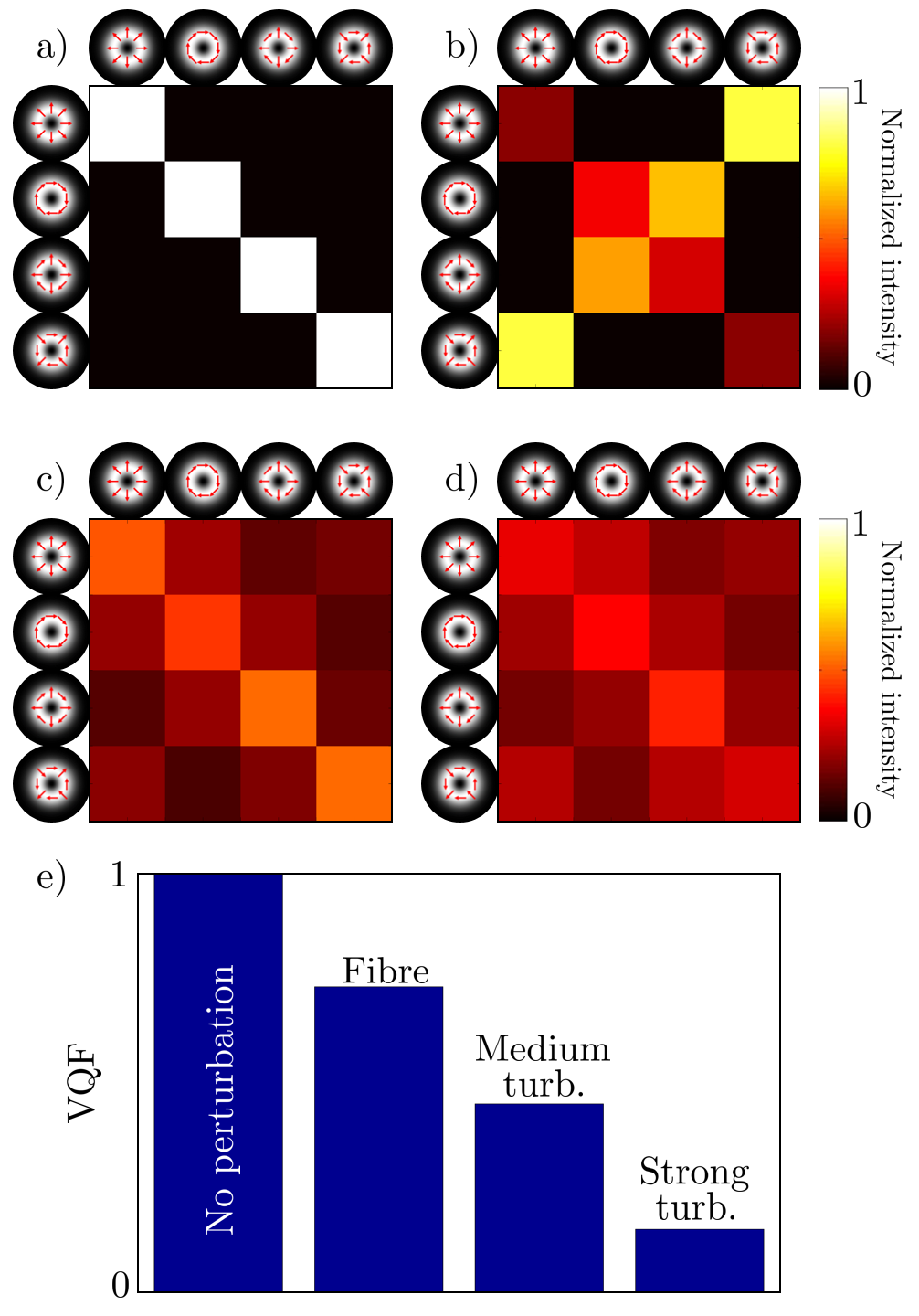}
	\caption{Crosstalk measurement of vector modes vector vortex modes propagating in (a) unperturbed conditions, (b) step index fibre, atmospheric turbulence at (c) medium ($SR = 0.6$) and (d) strong levels ($SR = 0.2$).  \AF{(e) As a result we note a decrease in the VQF from the initial idea conditions.}}
	\label{fig:figure9}
\end{figure}

We can confirm the above analysis experimentally.  In Fig.~\ref{fig:figure9}(a), we show the cross-talk matrix for a set of vector modes with $|\ell| = 1$, propagating in a non-perturbative medium. We then compare it to Fig.~\ref{fig:figure9}(b) which shows the the cross-talk matrix of the same vector modes propagating a 5 cm long step-index with core diameter 30 $\mu$m, supporting 76 modes at an operating wavelength of 633 nm.  Note that in general, one is more likely to measure a vector mode that is different from the one that was injected. This is because of the strong anti-diagonal nature of the cross-talk matrix that characterises the fibre. In Figs~\ref{fig:figure9}(c) and (d), the vector modes are made to propagated in medium and strong atmospheric turbulence conditions, respectively. The measure of turbulence used here is the Strehl ratio $SR$ \cite{Andrews2005}, ranging from 0 to 1 (0 is for very strong turbulence and 1 for no turbulence). Note that the detection probabilities diminish with increasing turbulence strength.  \AF{Following the characterisation method of the VQF, we find that as expected the VQF decreases due to the modal coupling, shown in Figs~\ref{fig:figure9}(e).} \BV{This decrease in VQF with increasing turbulence highlights the lack of resilience of vector vortex modes to atmospheric turbulence \cite{Cox2016}}.

\section{Conclusion}
In this work we have briefly reviewed the salient tools for the creation, characterisation and detection of vector vortex beams.  We have highlighted their applicability as information carriers in classical and quantum communication, and shown some examples of the degradation of such modes in perturbing media such as free-space with turbulence and imperfect optical fibre.  We present a comparison between probabilistic and deterministic detection schemes showing that the former, while ubiquitous, negates the very benefit of increased dimensionality in quantum communication while reducing signal in classical communication links.  Our newly proposed deterministic scheme overcomes this limitation.  We anticipate that this exciting field is likely to grow rapidly and believe that this work will offer a good starting point for new researchers in the field.

\end{document}